\documentclass[aps,12pt,showpacs,tightenlines,amsmath,amssymb]{revtex4}
\usepackage{graphicx}% Include figure files
\usepackage{dcolumn}% Align table columns on decimal point
\usepackage{bm}% bold math
\font\capsten=cmcsc10 scaled\magstep0

\begin{document}
%%%%%%%%%%%%%%%%%%%%%%%%%%%%%%%%%%%%%%%%%%%%%%%%%%%%%%%%%%%%%%%%%%%%%%
\title{Clusters from higher order correlations}
\author{L. S. Schulman}
\affiliation{Physics Department, Clarkson University, Potsdam, New York 13699-5820, USA}
\email{schulman@clarkson.edu}
\affiliation{Institut de Physique Th\'eorique CEA Saclay, Gif-sur-Yvette, F-91191, France}

\date{\today}
%%%%%%%%%%%%%%%%%%%%%%%%%%%%%%%%%%%%%%%%%%%%%%%%%%%%%%%%%%%%%%%%%%%%%%
\begin{abstract}
Given a set of variables and the correlations among them, we develop a method for finding clustering among the variables. The method takes advantage of information implicit in higher-order (not just pairwise) correlations. The idea is to define a Potts model whose energy is based on the correlations. Each state of this model is a partition of the variables and a Monte Carlo method is used to identify states of lowest energy, those most consistent with the correlations. A set of the 100 or so lowest such partitions is then used to construct a stochastic dynamics (using the adjacency matrix of each partition) whose observable representation gives the clustering. Three examples are studied. For two of them the 3$^\mathrm{rd}$ order correlations are significant for getting the clusters right. The last of these is a toy model of a biological system in which the joint action of several genes or proteins is necessary to accomplish a given process.
\end{abstract}
%%%%%%%%%%%%%%%%%%%%%%%%%%%%%%%%%%%%%%%%%%%%%%%%%%%%%%%%%%%%%%%%%%%%%%

\maketitle

Given a collection of objects that bear a relation to one another, it is often desirable to organize them into clusters or communities; moreover, this organization may take the form of a graph whose features not only allow the viewer a global image of a complex situation, but which may also have quantitative distance relations between and within the disparate communities. The applications are legion, and I mention several recent articles where this problem is discussed~\cite{blatt, grains, dynamicalmetric, ulanowicz, newman, girvan, ahn, ravasz, palla}.

There are also many kinds of relations among the ``objects,'' from flow (of probability, of substances, of attention), to collaboration, to correlation. In this article I focus on what can be learned from \textit{correlations} \cite{note:pca}, and especially from third- and higher-order correlation functions. As far as I know, no previous method has exploited this information. Such correlations may be expected to be particularly important in biological applications, where, for example, it often happens that it takes the concerted action of many genes or proteins to accomplish a given biological function.

The method makes use of previous work on the ``observable representation'' (OR) \cite{epigen, meanfieldobsrep, imaging}, but also introduces a Potts model calculation (see also \cite{blatt}). The principle of the method is clear enough, but the details can become complicated. We suppose that there are $N$ variables, $X_1,\dots,X_N$, and that they exhibit a number of correlations. Let each take $M$ values. As the measure of correlations we use normalized cumulants. Thus
 let $Y_k\equiv X_k-\langle X_k\rangle$, and $Z_k\equiv Y_k/\sqrt{\sum \langle Y_k^2\rangle/(M-1)}$, $k=1,\dots,N$. The quantities that will interest us are $J_{k\ell}\equiv\langle Z_k Z_\ell\rangle$, $K_{k\ell m}\equiv\langle Z_k Z_\ell Z_m\rangle$, etc. (for higher cumulants, additional terms will need to be subtracted from the expectation values).

The arrays $J$, $K$, etc., will be used as energy coefficients in a $q$ state Potts model. With appropriate sign, positively correlated variables will be in the same Potts state. I then use a stochastic dynamics based on this energy as a way of discovering the lowest energy configurations of this system. (Interestingly, the associated stochastic process has---up to a factor---as its \textit{high temperature} correlations just the original coefficients \cite{note:givepreciseversion}.) Each such configuration is essentially a \textit{partition} of the set of variables, $\{X\}$. A partition corresponds to an adjacency matrix and all such adjacency matrices are added with an appropriate Boltzmann weight. Finally, this $N$-by-$N$ matrix is treated as the generator of another stochastic process and its OR depicted.

\begin{figure}
\includegraphics[height=.325\textheight,width=.6\textwidth]{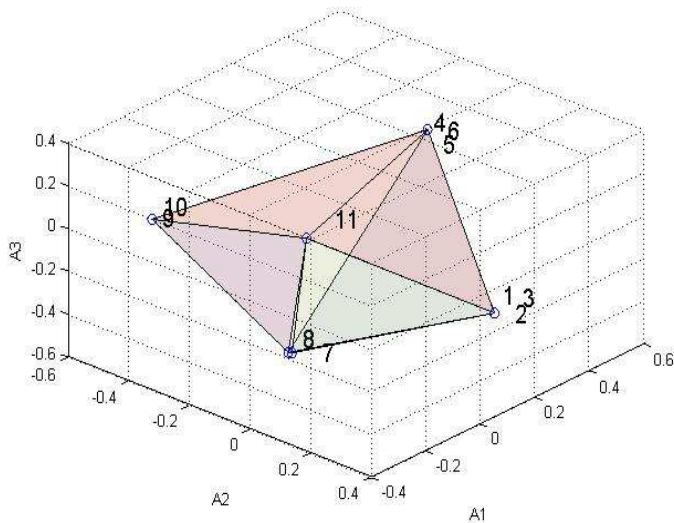}
\caption{Observable representation sorting of objects based on a fictitious dynamics in which the transition probabilities are (essentially) $\exp[-\sum_{ij}J_{ij}/T]$, with $J_{ij}$ the correlations. What is plotted is the triplet of left eigenvector values, $[A_1(x),A_2(x),A_3(x)]$, where the eigenvector labels correspond to descending algebraic value of the associated eigenvalue (the zeroth is the trivial $A_0(x)\equiv1$). Each little circle in the graph is one of the 11 points of the space and its identity printed near it. The convex hull of this set is indicated in light shading.}
\label{fig:ORJoverT}
\end{figure}

To illustrate the method, I work through three examples. In the first, there are 4 independent time series from which are built several others, in combinations that I will indicate. For this example, higher order correlations are not significant, so this will serve as an introduction to the general OR method. To obtain a stochastic matrix, the correlation functions, $J_{xy}$, are exponentiated (element-by-element) with a fictitious temperature, $T$. Thus, $R^{(0)}_{xy}=\exp[J_{xy}/T]$ for $x\ne y$. Let $\mu=\max_x\{\sum_{x, y\ne x} R^{(0)}_{xy}\}$ and let $R_{xy}= R^{(0)}_{xy}/\mu$ for $x\ne y$. Next, adjust the diagonal of $R$ so its column sums are unity \cite{note:easierinmatlab}. This yields a stochastic matrix ($R_{xy}$ is the probability of a transition, $y\to x$). Call the eigenvalues of $R$, $1=\lambda_0\ge\lambda_1\ge\dots$ (sometimes these are ordered by absolute value). The corresponding left and right eigenvectors are not (in general) equal, but since $J$ is symmetric, we have $p_\alpha(x)\propto A_\alpha(x)p_0(x)$, where the left eigenvectors are the $A$'s and the right eigenvectors the $p$'s. ($A_0\equiv1$, represents conservation of probability, and $p_0$ is the stationary state.) The temperature $T$ is adjusted rather loosely so as to bring $\lambda_1$ close to 1\@. In this example there are 11 variables. They are produced from 4 separate time series, with variables 1, 2 and 3 based on the first, 4, 5, and 6 on the second, 7 and 8 the third, 9 and 10 the fourth, while \#11 is mostly random, plus a small piece proportional to the sum of the generators of group 1 and group 2\@. The OR in Fig.\ \ref{fig:ORJoverT} is a 3-dimensional plot
of %    CORRECT THIS IN THE NEXT VERSION. WORD MISSING IN SUBMITTED MANUSCRIPT
the points $[A_1(x),A_2(x),A_3(x)]$ for $x$ in the space of 11 discrete variables. Note that the OR has grouped them properly. Since $\lambda_1$ is not extremely close to 1, some of the clusters are not at the extremal points. The 11$^\mathrm{th}$ point is kind of out of things, although had it had a smaller noise component it would have been on a line connecting group \#1 and group~\#2\@.

\begin{figure}
\includegraphics[height=.325\textheight,width=.6\textwidth]{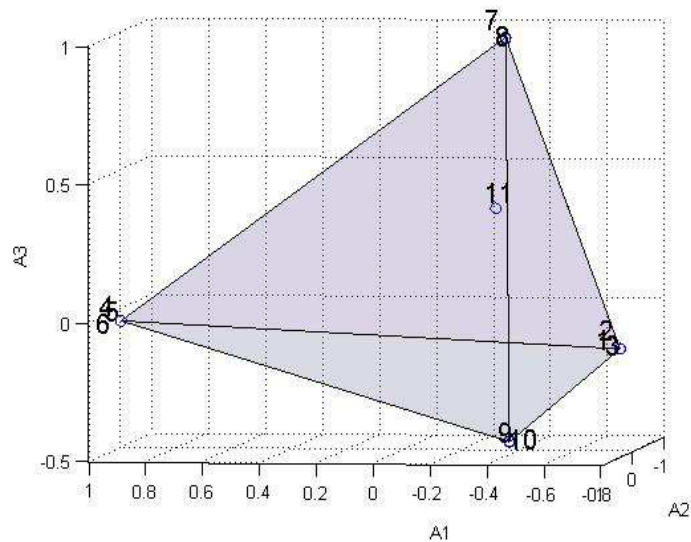}
\caption{Observable representation for a fictitious dynamics based on a weighted sum of adjacency matrices produced by a second fictitious stochastic process based on a Potts model whose energy is built from the correlation functions (including 3-point) of a set of variables. The convex hull of this set is indicated in light shading.}
\label{fig:OReig123}
\end{figure}

For the same set of variables, using both 2-point and 3-point correlations, Fig.\ \ref{fig:OReig123} shows the result of a lengthier process involving two stochastic processes. For a $q$-state Potts model let the energy be given by $E=-\sum \left(J_{k\ell}-J_0\right) \delta(s_k,s_\ell)- \sum_{k,\ell,m}K_{k\ell m} \delta(s_k,s_\ell,s_m)$, with $J_{k\ell}$ and $K_{k\ell m}$ correlations (as defined above) and $J_0$ a parameter, like $q$, to be selected based on considerations to be discussed in a moment. A Monte Carlo simulation is run for this system at a moderately low temperature, $T$. I have found that the most useful outcomes are obtained when, after an initial warmup stage, the system moves among the 100 or so lowest energy states, and $T$ is selected so that the system settles into this number states. (So far, missing sectors of state space, as could be expected in spin-glass-like structures, have not been in evidence; however, for some systems they could certainly appear.) These low energy states are degenerate under permutations of Potts-labels; however, each uniquely defines a partition of the original variables. The parameter $J_0$ is selected for non-trivial partitioning; thus it would avoid favoring only purely ferromagnetic states in the case where all correlations are positive. Finally, $q$ is selected large enough to give non-trivial partitions; taking it larger should not (and does not in practice, once a reasonable $q$ is selected) change the partitions.

Note in particular that this process uses three-point correlations, and can use yet higher correlations in an obvious way.

Once the significant partitions are identified an adjacency matrix is constructed for each. These are $N$-by-$N$ matrices (with $N$ the number of original variables). For each of these there is a frequency of occurrence in the simulation and an $N$-by-$N$ matrix is constructed by adding these adjacency matrices with that weight. Finally, this matrix is made into a stochastic matrix as in \cite{note:easierinmatlab} or by dividing by column sums (the former method preserves the symmetry of the correlations). Fig.\ \ref{fig:OReig123} is the OR for the first 3 (non-trivial) left eigenfunctions of that stochastic matrix.

It is evident, that both the direct use of $\exp(J/T)$ and the more complicated method using the Potts model have successfully divined the appropriate clustering of variables. I remark that in this example $K$ is quite small.

\begin{figure}
\centerline{\includegraphics[height=.2\textheight,width=.4\textwidth]{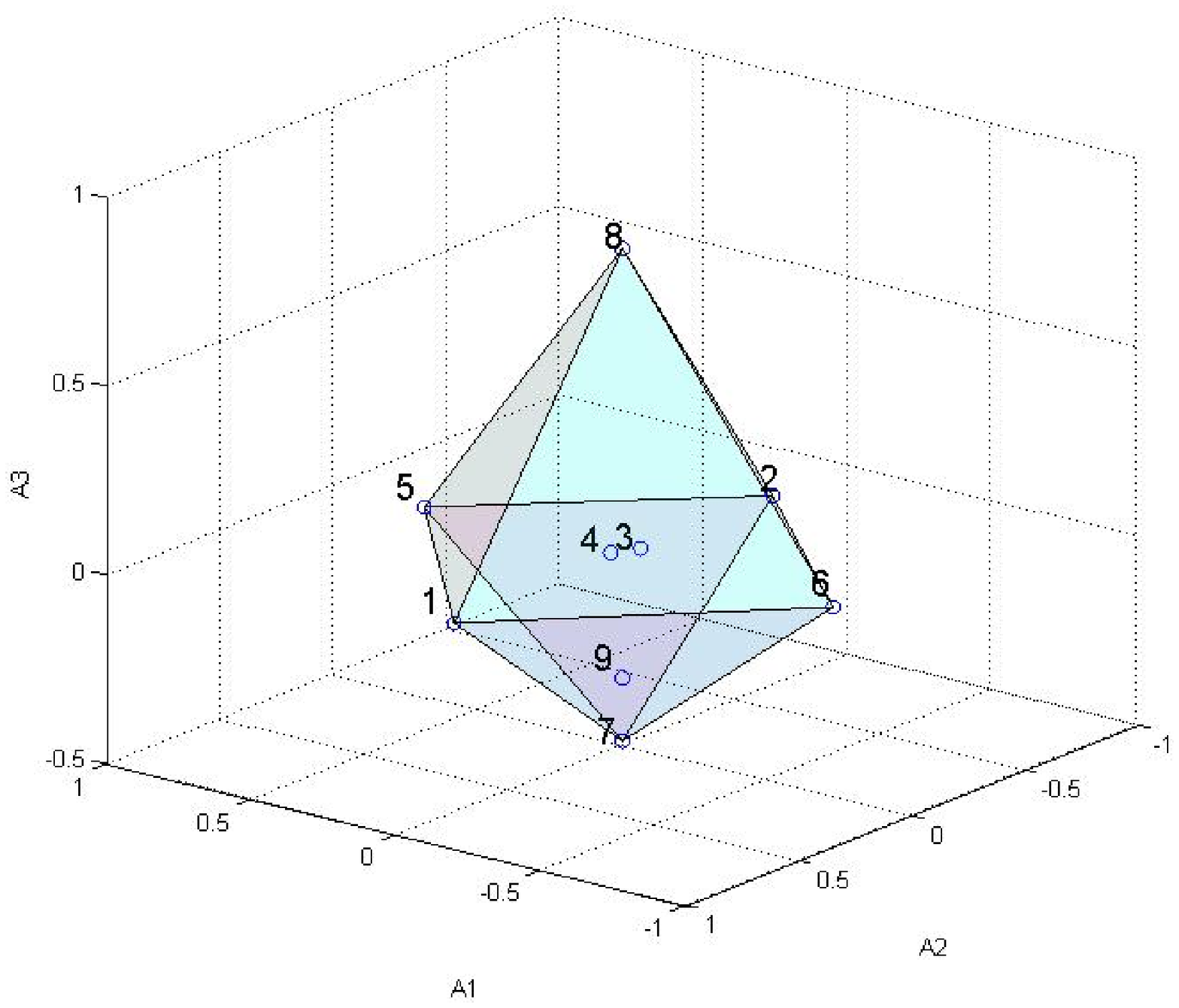}~
\includegraphics[height=.2\textheight,width=.4\textwidth]{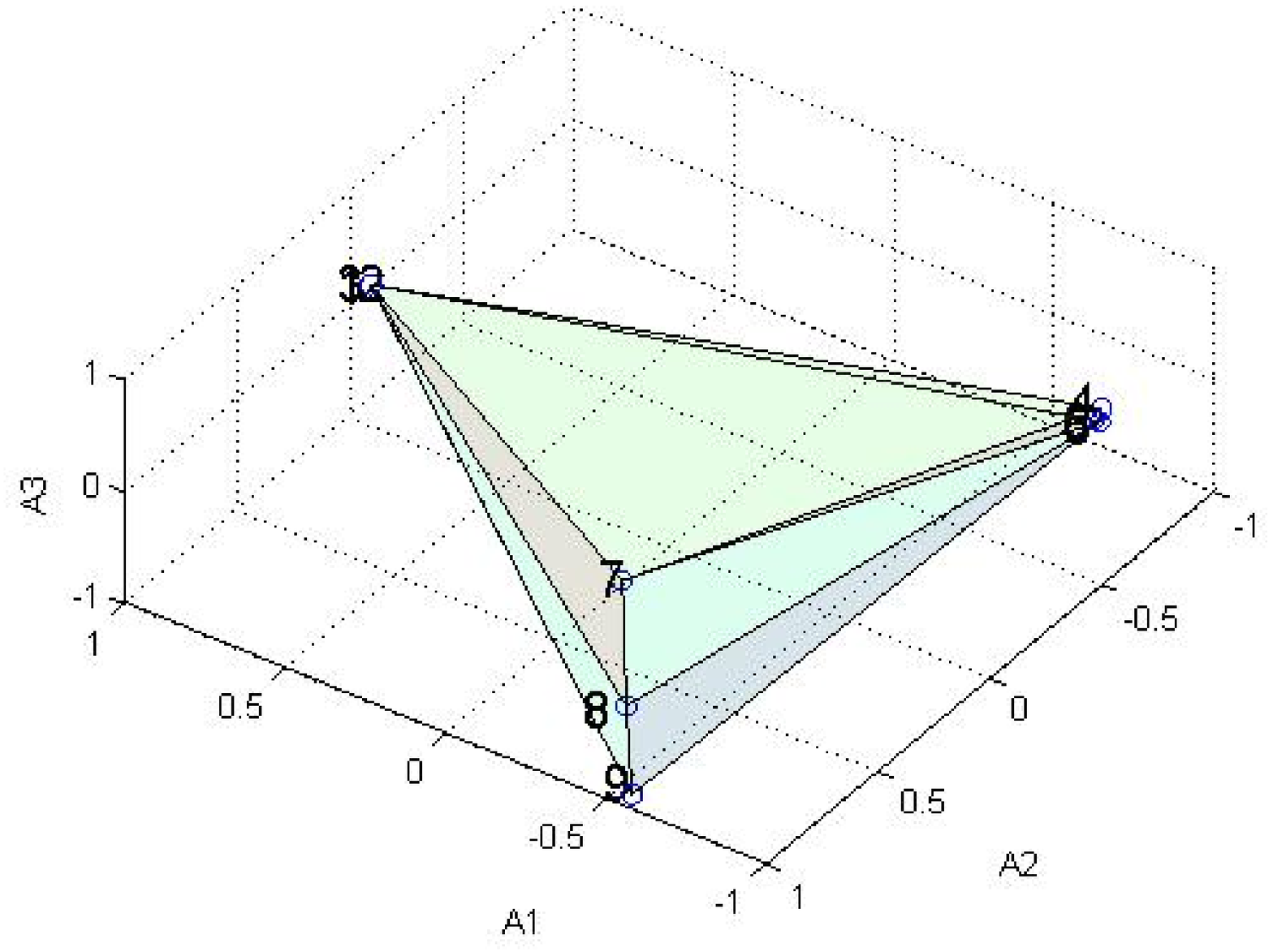}}
\caption{Observable representation. In the first figure, using $\exp(J/T)$, only 2-point correlations relate the r.v.'s to one another. In the second, using the Potts model/OR method, the (correct) grouping (1,2,3), (4,5,6) and (7,8,9) is found.}
\label{fig:Series2}
\end{figure}

A second example focuses on a case where the 3-point correlations are much larger than the 2-point correlations. It is well known that for r.v.'s $X$ and $Y$, independence implies $\langle XY\rangle=\langle X\rangle \langle Y\rangle$, but \textit{not} vice versa. By a Monte Carlo method \cite{note:triple} I produced three triples of r.v.'s with 2-point correlations at the level of 10$^{-4}$, and with 3-point correlations more than 100 times larger. As for Figs.\ \ref{fig:ORJoverT} and \ref{fig:OReig123}, I show in Fig.\ \ref{fig:Series2} the resulting clustering. The nine r.v.'s were grouped as 1-2-3, 4-5-6, and 7-8-9 and the Potts model method clearly displays that relation. On the other hand, based on 2-point functions and the $\exp(J/T)$ method, no relation among these variables is noted at all.

A third situation where the OR/Potts model method can be useful is where there may be several ``actors'' and ``processes.'' I have in mind a cartoon model of biological processes in which the actors (henceforth sans quotation marks) may be proteins or genes and the processes reactions that the actors may (collectively or individually) excite or inhibit. As I will illustrate, this can pinpoint actors that tend to work together or in opposition and may also be useful in discerning the ``processes.'' The information needed for this would be correlations (of 2$^\mathrm{nd}$ and higher order), something that for gene expression is now available \cite{epigen}.

I will now give an example in which the usual correlations give no information and one is \textit{only} able to discern an underlying grouping using 3$^\mathrm{rd}$ order correlations. I do not need to stress that this example is contrived to make exactly this point, and that for realistic situations one would get partial information at each level of correlations.

\begin{figure}
\centerline{\includegraphics[height=.2\textheight,width=.4\textwidth]{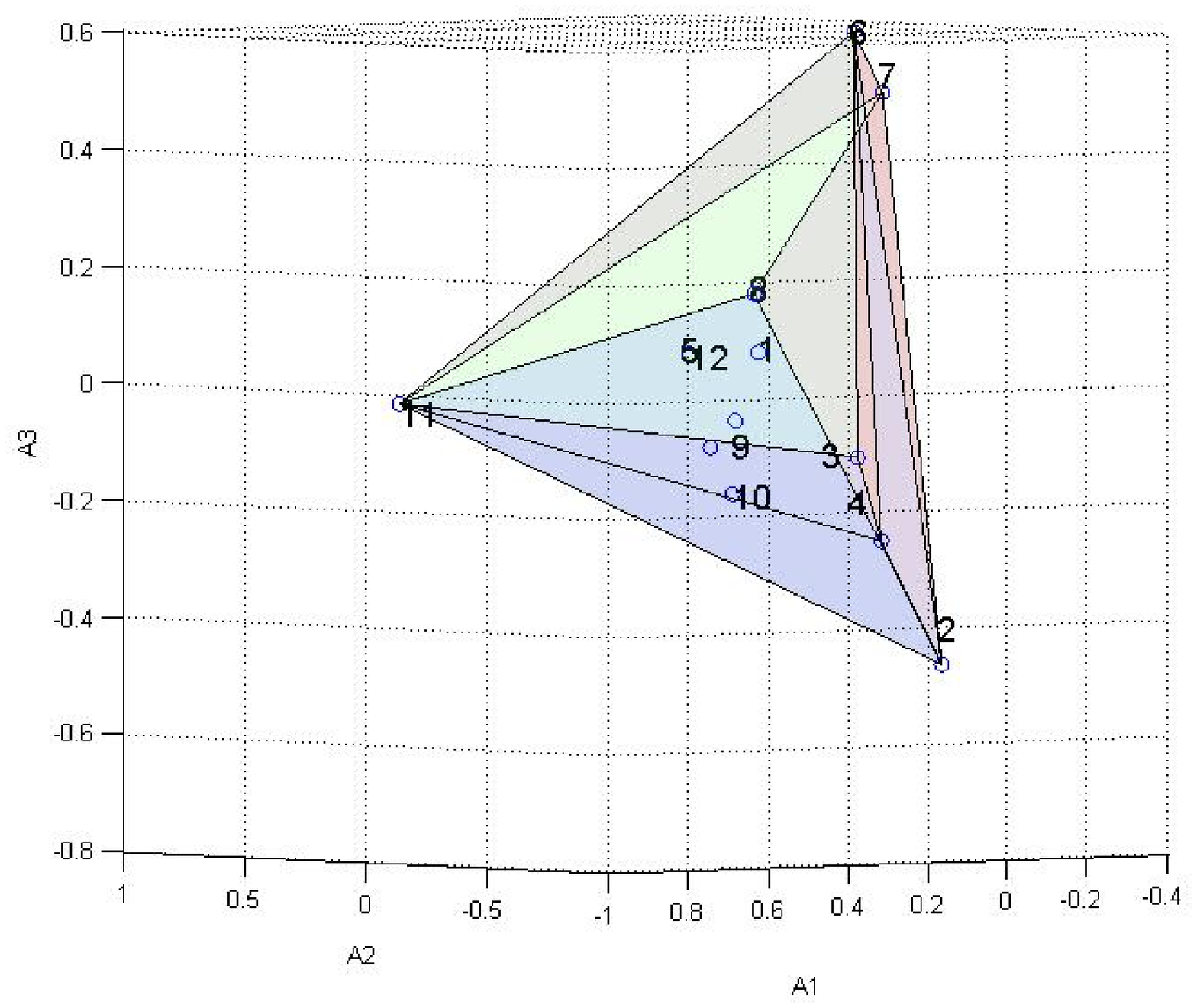}~
\includegraphics[height=.2\textheight,width=.4\textwidth]{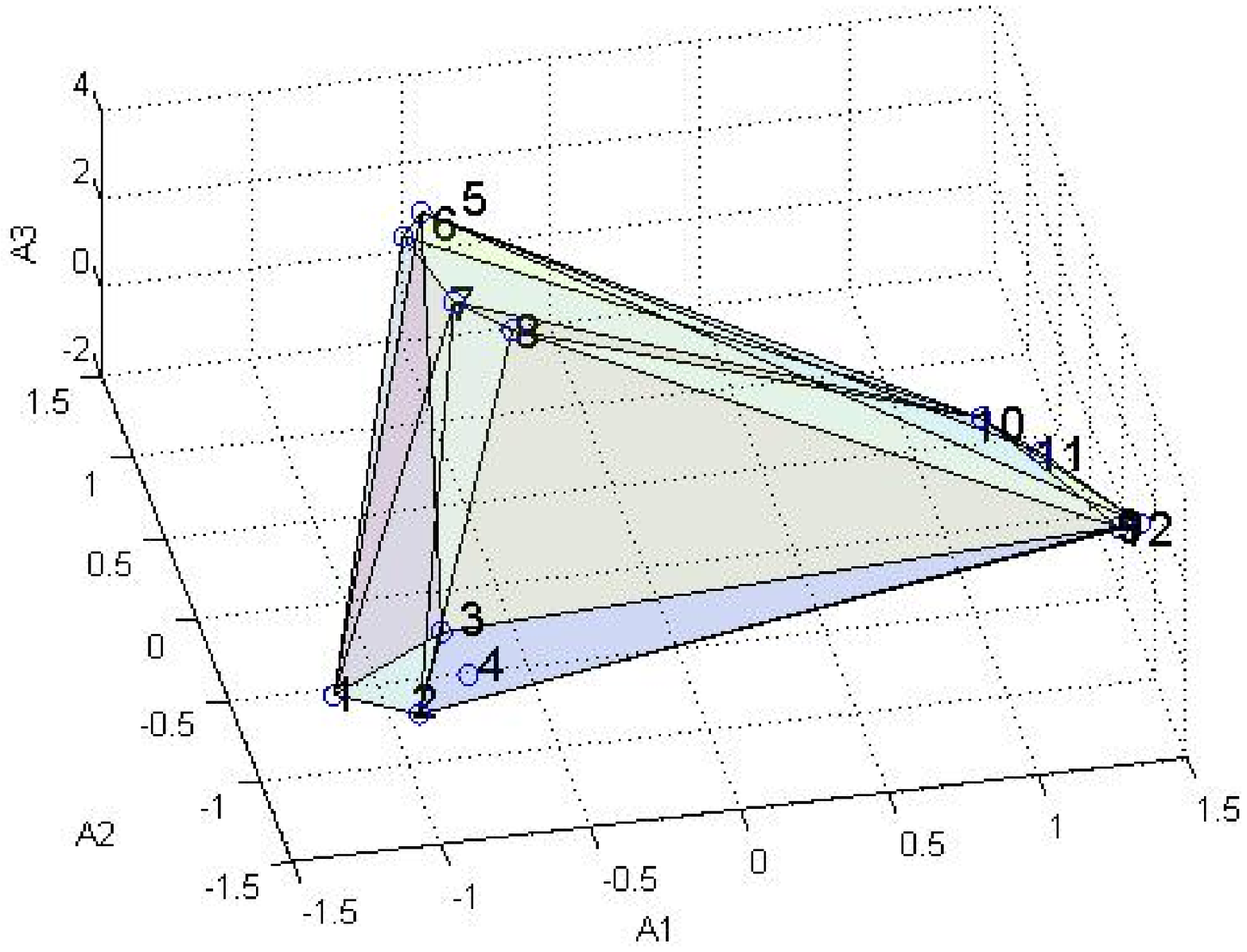}}
\caption{Observable representation. In the first figure only 2-point correlations have been used to relate the r.v.'s to one another. In the second, using the Potts model/OR method, the (correct) grouping (1,2,3,4), (5,6,7,8) and (9,10,11,12) is found. Whatever grouping may or may not be evident in the first figure is an artifact, since the eigenvalues are degenerate and the selection of which eigenvectors to plot is based on {\tiny{\capsten matlab}\texttrademark}'s arbitrary choices.}
\label{fig:actors}
\end{figure}

Suppose we have 12 actors and 59 processes. Three of the processes are performed by having (only) actors [1,2,3,4], [5,6,7,8] and [9,10,11,12] participating. The other 56 processes are all those that involve only 2 actors, omitting those pairs that appear in the first 3 processes. Thus I can write the process involving [1,2,3,4] as a vector: [1,1,1,1,0,0,0,0,0,0,0,0], with each column corresponding to an actor, and 1's indicating that the actor is involved in the process. With this notation, among the 56 other processes will be [1,0,0,0,1,0,0,0,0,0,0,0], involving \#1 and \#5. But [1,1,0,0,0,0,0,0,0,0,0,0] does not appear, since the pair [1,2] is included in one of the first three processes. For this pattern of activity, the two point correlations are essentially constant and the only information arises from 3 point functions. See Fig.~\ref{fig:actors}.

In conclusion I have shown how correlations, including (and especially) 3$^\mathrm{rd}$ order and higher, can be used to group variables. Calling the 2-point correlations, $J$, a simple study of the matrix $\exp(J_{xy}/T)$ (with $T$ a temperature-like parameter) can already give such information using the observable representation (OR) and by many other methods. However, to incorporate information available from higher order correlations, we go to an exponentially larger space on which we consider a random process based on a Potts model whose energy is related to all the correlations. For this process the OR is impractical (because of the size of the space), but the usual kind of Monte Carlo study can identify the lowest energy states. Using these we deduce the most significant partitions of the original variables and use the OR on the original space to deduce groupings.

%%%%%%%%%%%%%%%%%%%%%%%%%%%%%%%%%%%%%%%%%%%%%%%%%%%%%%%%%%%%%%%%%%%%%%%%%%%
\begin{acknowledgments}
I thank Bernard Gaveau and Leonard Schulman for helpful discussions. This work was supported by the United States National Science Foundation Grant PHY~05~55313.
\end{acknowledgments}
%%%%%%%%%%%%%%%%%%%%%%%%%%%%%%%%%%%%%%%%%%%%%%%%%%%%%%%%%%%%%%%%%%%%%%%%%%%

%\bibliography{ClusterCorrelations,c:/LS/Articles/InPrep/LSSComp09}

\begin{thebibliography}{17}
\expandafter\ifx\csname natexlab\endcsname\relax\def\natexlab#1{#1}\fi
\expandafter\ifx\csname bibnamefont\endcsname\relax
  \def\bibnamefont#1{#1}\fi
\expandafter\ifx\csname bibfnamefont\endcsname\relax
  \def\bibfnamefont#1{#1}\fi
\expandafter\ifx\csname citenamefont\endcsname\relax
  \def\citenamefont#1{#1}\fi
\expandafter\ifx\csname url\endcsname\relax
  \def\url#1{\texttt{#1}}\fi
\expandafter\ifx\csname urlprefix\endcsname\relax\def\urlprefix{URL }\fi
\providecommand{\bibinfo}[2]{#2}
\providecommand{\eprint}[2][]{\url{#2}}

\bibitem[{\citenamefont{Blatt et~al.}(1996)\citenamefont{Blatt, Wiseman, and
  Domany}}]{blatt}
\bibinfo{author}{\bibfnamefont{M.}~\bibnamefont{Blatt}},
  \bibinfo{author}{\bibfnamefont{S.}~\bibnamefont{Wiseman}}, \bibnamefont{and}
  \bibinfo{author}{\bibfnamefont{E.}~\bibnamefont{Domany}},
  \emph{\bibinfo{title}{Superparamagnetic Clustering of Data}},
  \bibinfo{journal}{Phys.\ Rev.\ Lett.} \textbf{\bibinfo{volume}{76}},
  \bibinfo{pages}{3251-3254} (\bibinfo{year}{1996}). \bibinfo{note}{This work also introduces a Potts model, but goes in a
  direction opposite to ours, \textit{from} distances, \textit{to} correlations
  (and clusters). See also~\cite{reichardt}}.

\bibitem[{\citenamefont{Schulman and Gaveau}(2001)}]{grains}
\bibinfo{author}{\bibfnamefont{L.~S.} \bibnamefont{Schulman}} \bibnamefont{and}
  \bibinfo{author}{\bibfnamefont{B.}~\bibnamefont{Gaveau}},
  \emph{\bibinfo{title}{Coarse grains: the emergence of space and order}},
  \bibinfo{journal}{Found.\ Phys.} \textbf{\bibinfo{volume}{31}},
  \bibinfo{pages}{713-731} (\bibinfo{year}{2001}).

\bibitem[{\citenamefont{Gaveau and Schulman}(2005)}]{dynamicalmetric}
\bibinfo{author}{\bibfnamefont{B.}~\bibnamefont{Gaveau}} \bibnamefont{and}
  \bibinfo{author}{\bibfnamefont{L.~S.} \bibnamefont{Schulman}},
  \emph{\bibinfo{title}{Dynamical distance: coarse grains, pattern recognition,
  and network analysis}}, \bibinfo{journal}{Bull.\ Sci.\ math.}
  \textbf{\bibinfo{volume}{129}}, \bibinfo{pages}{631-642}
  (\bibinfo{year}{2005}).

\bibitem[{\citenamefont{Ulanowicz}(2004)}]{ulanowicz}
\bibinfo{author}{\bibfnamefont{R.~E.} \bibnamefont{Ulanowicz}},
  \emph{\bibinfo{title}{Quantitative methods for ecological network analysis}},
  \bibinfo{journal}{Comp. Bio. Chem.} \textbf{\bibinfo{volume}{28}},
  \bibinfo{pages}{321-339} (\bibinfo{year}{2004}).

\bibitem[{\citenamefont{Newman et~al.}(2002)\citenamefont{Newman, Watts, and
  Strogatz}}]{newman}
\bibinfo{author}{\bibfnamefont{M.~E.~J.} \bibnamefont{Newman}},
  \bibinfo{author}{\bibfnamefont{D.~J.} \bibnamefont{Watts}}, \bibnamefont{and}
  \bibinfo{author}{\bibfnamefont{S.~H.} \bibnamefont{Strogatz}},
  \emph{\bibinfo{title}{Random graph models of social networks}},
  \bibinfo{journal}{Proc. Natl. Acad. Sci.} \textbf{\bibinfo{volume}{99}},
  \bibinfo{pages}{2566-2572} (\bibinfo{year}{2002}).

\bibitem[{\citenamefont{Girvan and Newman}(2002)}]{girvan}
\bibinfo{author}{\bibfnamefont{M.}~\bibnamefont{Girvan}} \bibnamefont{and}
  \bibinfo{author}{\bibfnamefont{M.~E.~J.} \bibnamefont{Newman}},
  \emph{\bibinfo{title}{Community structure in social and biological
  networks}}, \bibinfo{journal}{Proc. Natl. Acad. Sci.}
  \textbf{\bibinfo{volume}{99}}, \bibinfo{pages}{7821-7826}
  (\bibinfo{year}{2002}).

\bibitem[{\citenamefont{Ahn et~al.}(2009)\citenamefont{Ahn, Bagrow, and
  Lehmann}}]{ahn}
\bibinfo{author}{\bibfnamefont{Y.-Y.} \bibnamefont{Ahn}},
  \bibinfo{author}{\bibfnamefont{J.}~\bibnamefont{Bagrow}}, \bibnamefont{and}
  \bibinfo{author}{\bibfnamefont{S.}~\bibnamefont{Lehmann}},
  \emph{\bibinfo{title}{Communities and Hierarchical Organization of Links in
  Complex Networks}}, \bibinfo{journal}{preprint}  (\bibinfo{year}{2009}),
  \bibinfo{note}{arXiv:0903.3178}.

\bibitem[{\citenamefont{Ravasz et~al.}(2002)\citenamefont{Ravasz, Somera,
  Mongru, Oltvai, and Barabasi}}]{ravasz}
\bibinfo{author}{\bibfnamefont{E.}~\bibnamefont{Ravasz}},
  \bibinfo{author}{\bibfnamefont{A.~L.} \bibnamefont{Somera}},
  \bibinfo{author}{\bibfnamefont{D.~A.} \bibnamefont{Mongru}},
  \bibinfo{author}{\bibfnamefont{Z.~N.} \bibnamefont{Oltvai}},
  \bibnamefont{and} \bibinfo{author}{\bibfnamefont{A.-L.}
  \bibnamefont{Barabasi}}, \emph{\bibinfo{title}{Hierarchical Organization of
  Modularity in Metabolic Networks}}, \bibinfo{journal}{Science}
  \textbf{\bibinfo{volume}{297}}, \bibinfo{pages}{1551-1555}
  (\bibinfo{year}{2002}).

\bibitem[{\citenamefont{Palla et~al.}(2005)\citenamefont{Palla, Der\'enyi,
  Farkas, and Vicsek}}]{palla}
\bibinfo{author}{\bibfnamefont{G.}~\bibnamefont{Palla}},
  \bibinfo{author}{\bibfnamefont{I.}~\bibnamefont{Der\'enyi}},
  \bibinfo{author}{\bibfnamefont{I.}~\bibnamefont{Farkas}}, \bibnamefont{and}
  \bibinfo{author}{\bibfnamefont{T.}~\bibnamefont{Vicsek}},
  \emph{\bibinfo{title}{Uncovering the overlapping community structure of
  complex networks in nature and society}}, \bibinfo{journal}{Nature}
  \textbf{\bibinfo{volume}{435}}, \bibinfo{pages}{814-818}
  (\bibinfo{year}{2005}).

\bibitem[{not({\natexlab{a}})}]{note:pca}
\bibinfo{note}{Information about correlations is often exploited using
  principal component analysis (PCA). A multidimensional plot of the
  eigenvectors of the correlation matrix ($J$) can accomplish some of the same
  objectives as our OR study of $\exp(J/T)$ (for parameter $T$). Neither $J$
  nor its exponentiation of course takes into account higher order
  correlations.}

\bibitem[{\citenamefont{Schulman et~al.}(2009)\citenamefont{Schulman, Bagrow,
  and Gaveau}}]{epigen}
\bibinfo{author}{\bibfnamefont{L.~S.} \bibnamefont{Schulman}},
  \bibinfo{author}{\bibfnamefont{J.~P.} \bibnamefont{Bagrow}},
  \bibnamefont{and} \bibinfo{author}{\bibfnamefont{B.}~\bibnamefont{Gaveau}},
  \emph{\bibinfo{title}{User's manual for the observable representation}}, in
  \emph{\bibinfo{booktitle}{Systems Epigenomics}}, edited by
  \bibinfo{editor}{\bibfnamefont{A.}~\bibnamefont{Benecke}}
  (\bibinfo{publisher}{Springer-Verlag}, \bibinfo{address}{Berlin},
  \bibinfo{year}{2009}), Springer Biology Series, \bibinfo{note}{to appear}.

\bibitem[{\citenamefont{Schulman}(2007)}]{meanfieldobsrep}
\bibinfo{author}{\bibfnamefont{L.~S.} \bibnamefont{Schulman}},
  \emph{\bibinfo{title}{Mean Field Spin Glass in the Observable
  Representation}}, \bibinfo{journal}{Phys.\ Rev.\ Lett.}
  \textbf{\bibinfo{volume}{98}}, \bibinfo{pages}{257202}
  (\bibinfo{year}{2007}).

\bibitem[{\citenamefont{Gaveau et~al.}(2006)\citenamefont{Gaveau, Schulman, and
  Schulman}}]{imaging}
\bibinfo{author}{\bibfnamefont{B.}~\bibnamefont{Gaveau}},
  \bibinfo{author}{\bibfnamefont{L.~S.} \bibnamefont{Schulman}},
  \bibnamefont{and} \bibinfo{author}{\bibfnamefont{L.~J.}
  \bibnamefont{Schulman}}, \emph{\bibinfo{title}{Imaging geometry through
  dynamics: the observable representation}}, \bibinfo{journal}{J. Phys.\ A}
  \textbf{\bibinfo{volume}{39}}, \bibinfo{pages}{10307-10321}
  (\bibinfo{year}{2006}).

\bibitem[{not({\natexlab{b}})}]{note:givepreciseversion}
\bibinfo{note}{% See my notes of 11/7/08 For the $q$-state Potts model,
  $\langle \bar\delta(s_i,s_j)\rangle\to %\frac{q-1}{q^2T}
  \left((q-1)/q^2\right) J_{ij}$ for $T\to\infty$. Here $\delta$ is the
  Kronecker delta, $s_i$ the value (1 to $q$) taken by the Potts variable by
  variable-$i$, $T$ the temperature, and
  $\bar\delta\equiv\delta-\langle\delta\rangle$. The mean value,
  $\langle\delta\rangle$ is $1/q$. For higher-order correlations similar limits
  hold. These properties are not difficult to show.}

\bibitem[{not({\natexlab{c}})}]{note:easierinmatlab}
\bibinfo{note}{This prescription is more easily stated in {\tiny{\capsten
  matlab}\texttrademark} notation: R = exp(J/T); R = R $-$ diag(diag(R)); M =
  max(sum(R)); R = R/M; R = eye(size(R)) $-$ diag(sum(R)) $+$ R;}.

\bibitem[{not({\natexlab{d}})}]{note:triple}
\bibinfo{note}{If each of 3 r.v.'s takes $N$ values, the probabiity
  distribution, $p$, consists of $N^3$ non-negative numbers adding to 1\@. I
  did a Monte Carlo run on these numbers, first aiming to lower all 2-point
  correlations. Then, holding the sum of the absolute values of the 2-point
  functions to not exceed the value attained, I continued the Monte Carlo run
  in order to increase the 3-point functions.}

\bibitem[{\citenamefont{Reichardt and Bornholdt}(2004)}]{reichardt}
\bibinfo{author}{\bibfnamefont{J.}~\bibnamefont{Reichardt}} \bibnamefont{and}
  \bibinfo{author}{\bibfnamefont{S.}~\bibnamefont{Bornholdt}},
  \emph{\bibinfo{title}{Detecting fuzzy community structures in complex
  networks with a q-state Potts model}}, \bibinfo{journal}{Phys.\ Rev.\ Lett.}
  \textbf{\bibinfo{volume}{93}}, \bibinfo{pages}{218701}
  (\bibinfo{year}{2004}).

\end{thebibliography}

\end{document}